\documentclass[manuscript]{aastex}

\shorttitle{Synthesis of molecular oxygen via irradiation of ice grains in the protosolar nebula}
\shortauthors{Mousis et al.}

\usepackage{etex}
\usepackage{m-pictex, m-ch-en}
\usepackage {amsmath}
\usepackage[squaren, Gray, cdot]{SIunits}
\usepackage{color}

\begin{document}


\title{Synthesis of molecular oxygen via irradiation of ice grains in the protosolar nebula}


\author{O. Mousis\altaffilmark{1}, T. Ronnet\altaffilmark{1}, J. I. Lunine\altaffilmark{2}, R. Maggiolo\altaffilmark{3}, P. Wurz\altaffilmark{4}, G. Danger\altaffilmark{5} and A. Bouquet\altaffilmark{6}}


\altaffiltext{1}{Aix Marseille Universit{\'e}, CNRS, LAM (Laboratoire d'Astrophysique de Marseille) UMR 7326, 13388, Marseille, France {\tt olivier.mousis@lam.fr}}
\altaffiltext{2}{Department of Astronomy, Cornell University, Ithaca, NY 14853, USA}
\altaffiltext{3}{Royal Belgian Institute for Space Aeronomy, BIRA-IASB, Ringlaan 3, B-1180 Brussels, Belgium}
\altaffiltext{4}{Physikalisches Institut, University of Bern, Sidlerstrasse 5, CH-3012 Bern, Switzerland}
\altaffiltext{5}{Aix-Marseille Universit\'e, PIIM UMR-CNRS 7345, F-13397 Marseille, France}
\altaffiltext{6}{Department of Space Research, Southwest Research Institute, 6220 Culebra Rd., San Antonio, TX 78228, USA}


\begin{abstract}
Molecular oxygen has been detected in the coma of comet 67P/Churyumov--Gerasimenko with a mean abundance of 3.80 $\pm$ 0.85\% by the ROSINA mass spectrometer on board the Rosetta spacecraft. To account for the presence of this species in comet 67P/Churyumov--Gerasimenko, it has been shown that the radiolysis of ice grains precursors of comets is a viable mechanism in low-density environments, such as molecular clouds. Here, we investigate the alternative possibility that the icy grains present in the midplane of the protosolar nebula were irradiated during their vertical transport between the midplane and the upper layers over a large number of cycles, as a result of turbulent mixing. Consequently, these grains spent a non-negligible fraction of their lifetime in the disk's upper regions, where the irradiation by cosmic rays was strong. To do so, we used a coupled disk-transport-irradiation model to calculate the time evolution of the molecular oxygen abundance radiolytically produced in ice grains. Our computations show that, even if a significant fraction of the icy particles have followed a back and forth cycle towards the upper layers of the disk over 10 million of years, a timespan far exceeding the formation timescale of comet 67P/Churyumov--Gerasimenko, the amount of produced molecular oxygen is at least two orders of magnitude lower than the Rosetta observations. We conclude that the most likely scenario remains the formation of molecular oxygen in low-density environments, such as the presolar cloud, prior to the genesis of the protosolar nebula.
\end{abstract}

\keywords{comets: general -- comets: individual (67P/Churyumov-Gerasimenko) -- solid state: volatile -- methods: numerical -- astrobiology}

\section{Introduction}

Molecular oxygen (O$_2$) has been detected in the coma of comet 67P/Churyumov--Gerasimenko (67P/C-G) with abundances in the 1--10\% range and a mean value of 3.80 $\pm$ 0.85\% by the Rosetta Orbiter Spectrometer for Ion and Neutral Analysis-Double Focusing Mass Spectrometer (ROSINA) instrument on board the Rosetta spacecraft \citep{Bi15}. Moreover, the production rate of O$_2$ has been found remarkably correlated with that of H$_2$O in 67P/C-G's coma, suggesting that both molecules come from the same icy phase \citep{Bi15}. A subsequent reanalysis of the Giotto mass spectrometry data shows that O$_2$ was also present in the coma of comet 1P/Halley with an abundance of 3.7~$\pm$~1.7\% with respect to water at the time of its encounter with the ESA spacecraft, suggesting that this species could be a common parent species in comets \citep{Ru15}.
 
To account for the O$_2$ abundance and the correlation of its production rate with that of H$_2$O in 67P/C-G, it has been shown that the radiolysis of ice grains in low-density environments, such as molecular clouds, is a mechanism capable of producing large amounts of O$_2$ from solid H$_2$O \citep{Mo16}. Meanwhile, it was also found that the radiolysis of icy grains in higher density environments, such as the midplane of protoplanetary disks, is not efficient enough to create amounts of O$_2$ comparable with those observed in 67P/C-G and 1P/Halley in timescales shorter than the lifetime of the protosolar nebula, even in the case of a strong cosmic ray flux (CRF) engendered by the presence of a nearby supernova \citep{Mo16}. It was then concluded that, if O$_2$ is an irradiation product of ice, this molecule was probably formed in the interstellar medium, prior to the formation of the PSN.

\begin{figure}[h]
\centerline{\includegraphics[width=\columnwidth]{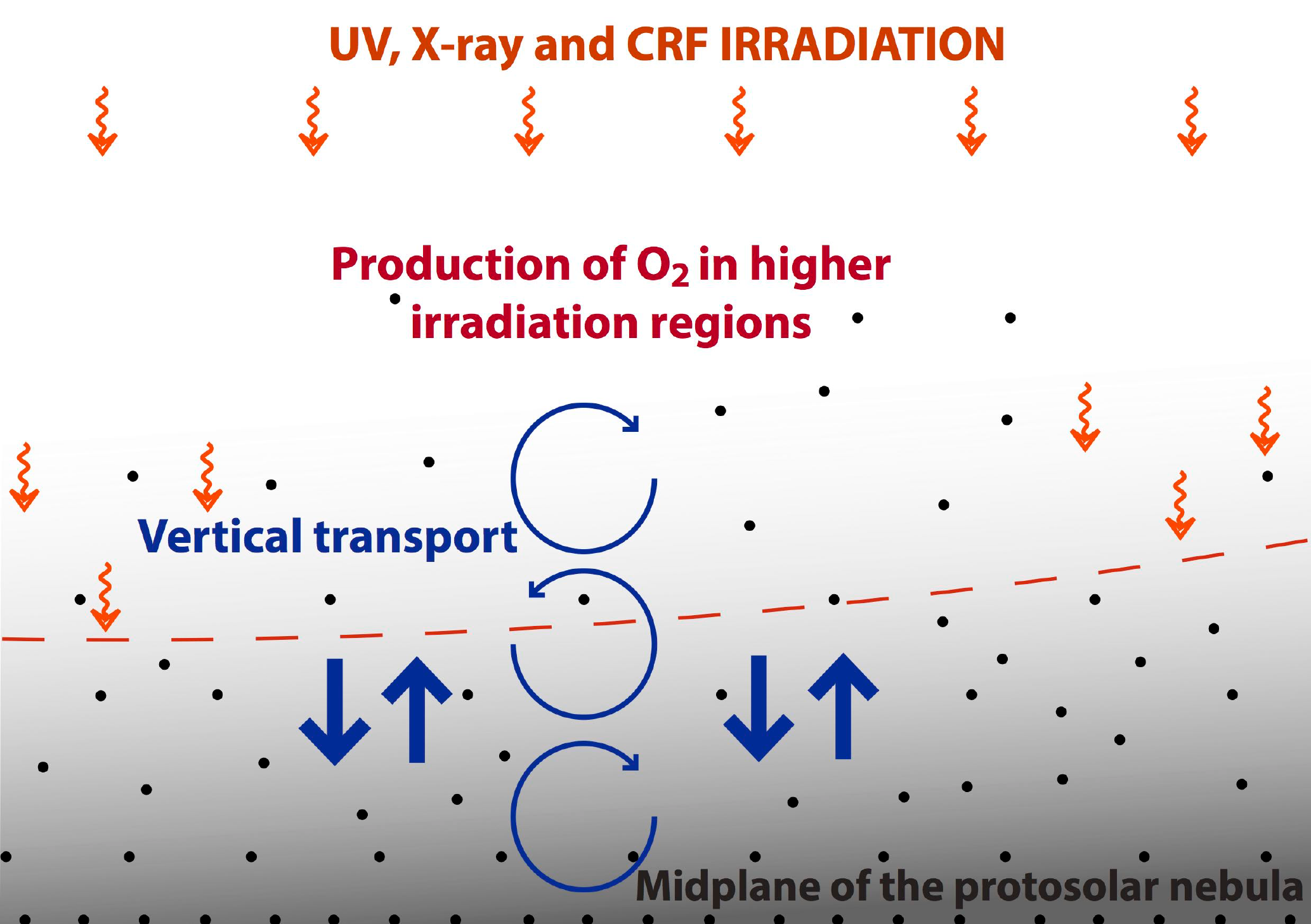}}
\caption{Illustration of the vertical transport of small icy grains toward disk regions where irradiation is strong and favors the formation of O$_2$. Grains remain concentrated in the midplane of the disk because of gravitational settling and gas drag.}
\label{fig1}
\end{figure}

Here, we consider the alternative possibility that the icy grains present in the midplane of the PSN were irradiated during their transport towards its upper layers (see illustration in Figure \ref{fig1}). This idea has been introduced in \cite{Mo17a} but never investigated to quantitatively estimate the fraction of O$_2$ produced during the vertical transport of grains. Due to turbulent mixing, the ice grains present in the midplane were lifted toward the upper layers of the disk and dragged down over a large number of cycles \citep{Ci12,Mo17a}. Consequently, these grains spent a non-negligible fraction of their lifetime in the disk's upper regions, where the irradiation by cosmic rays was strong. This irradiation, integrated over the disk lifetime , might cause some potentially significant production of O$_2$, the extent of which is explored in the present work.

\section{Disk model and transport module}

To mimic the vertical motion of particles, we used a simple description of the PSN structure \citep{Ch97,Ha98,Ci12}. The gas surface density $\Sigma_g$, {\bf namely the gas column density integrated through the disk plane}, and temperature $T_d$ profiles of the disk model are given by: 

\begin{equation}
\Sigma_g (r) = 2000 \left(\frac{r}{1 AU} \right)^{-1} {\rm g~cm^{-2}}
\end{equation}

\noindent and

\begin{equation}
T_d (r) = 280 \left(\frac{r}{1 AU} \right)^{-1/2} {\rm K},
\end{equation}
  
\noindent where $r$ is the heliocentric distance. These prescriptions are appropriate for flared disks around young T Tauri stars. 

Using the assumption of vertical hydrostatic equilibrium for the gas, the disk's density is expressed as a function of heliocentric distance $r$ and altitude $z$ above the midplane as:

\begin{equation}
\rho_g(r,z) = \rho_0(r) e^{-\frac{z^2}{2H^2_g}},
\end{equation}

\noindent with

\begin{equation}
\rho_0(r) =  \frac{\Sigma_g(r)}{\sqrt{2\pi} H_g},
\end{equation}

\noindent where $H_g = c_g / \Omega_K$ is the gas scale height derived from the isothermal gas sound speed $c_g = \sqrt{R_g T_d / \mu_g}$ and the keplerian orbital frequency $\Omega_K$. $R_g$ is the ideal gas constant and $\mu_g$ the mean molecular weight of the gas ($\sim$2.4 g/mol).

Below we briefly outline the main aspects of the transport model used in our computations. We refer the reader to the work of \cite{RMV17} for a full description. We consider the diffusion of dust grains onto the background disk gas using a Lagrangian approach where individual grains are tracked. Because our study only addresses the influence of irradiation during the vertical transport of grains, we opted not to follow their radial evolution. Small dust grains are strongly coupled and have a radial velocity similar to that of the background gas (and follow the accretion flow onto the star), whereas larger, partially decoupled grains, rapidly drift inward due to gas drag \citep[e.g.][]{We77}. Considering the vertical hydrostatic equilibrium of the gas disk, we assume the gas has no net vertical velocity \citep[see][for a discussion of this assumption]{TL02}.

The vertical motion of the grains is therefore ruled by the turbulent diffusion that lifts them towards the upper layers of the disk and the settling towards the midplane due to gas drag and the gravity of the central star. {\bf The latter} is given by the equation of motion of the dust grains in the vertical direction, which is solved following the approach developed in \citet{RMV17}:

\begin{equation} 
\label{eqmo}
\frac{d v_{d,z}}{dt} = -\frac{GMz}{r^3} - \frac{v_{d,z}}{t_s},
\end{equation}

\noindent where $v_{d,z}$ is the dust vertical velocity, $M=1\,M_\odot$ is the mass of the central star, $z$ is the vertical position of the dust grain, and $t_s$ its the stopping time. The stopping time is a measure of the timescale on which the gas transfers its angular momentum to dust and can be expressed as \citep{PM11}:

\begin{equation}
t_s = \left( \frac{\rho_g v_{th}}{\rho_s R_s} \mathrm{min}\left[ 1, \frac{3}{8} \frac{v_{rel}}{v_{th}} C_D(Re)\right] \right)^{-1},
\end{equation}

\noindent where $v_{th} = \sqrt{8/\pi}c_g$ is the gas thermal velocity, $R_s$ is the radius of the solid particle, $\rho_s$ is its density, assumed to be 1 g cm$^{-3}$ regardless of the size, and $v_{rel}$ is the relative velocity between the gas and the dust grain. The dimensionless drag coefficient $C_D$ is a function of the Reynolds number $Re$ of the flow around the particle and derives from an empirical law fitted on recent experimental data \citep{PM11}:
 
\begin{equation}
C_D=\frac{24}{Re} (1+0.27Re)^{0.43}+0.47\left(1-e^{-0.04Re^{0.38}}\right),
\end{equation} 
 
\noindent where $Re$ is the Reynolds number. It is given by \citep{SL00}:

\begin{equation}
Re = \frac{4 R_s v_{rel}}{c_g l_g},
\end{equation}

\noindent where $l_g$ is the mean free path of the gas.\\

The turbulent diffusion of the grains is modeled using a Monte-Carlo scheme where individual particles are given random impulses to mimic the stochastic transport due to turbulent eddies \citep[see e.g.,][]{Ci10,Ci11}. Overall, the new position $z_d$ of a dust grain after a timestep $dt$ is computed as follows in the vertical direction:

\begin{equation}
z_d(t+dt) = z_d(t) + v_{adv} dt + R_1 \left[ \frac{2}{\sigma^2} D_p dt \right]^{\frac{1}{2}},
\end{equation}

\noindent where $R_1 \in [-1;1]$ is a random number, $\sigma^2$ the variance of the random number distribution, $D_p$ the diffusivity of the solid particle and $v_{adv}$ is the term accounting for the non-uniform density of the gas in which the particles diffuse as well as the non uniform diffusivity of the particles, and the forces experienced by the particle (see eq.~\ref{adv}). The diffusivity of the dust grains $D_p$ is related to the gas diffusivity through the Schmidt number $Sc$ as \citep{YL07}:

\begin{equation}
Sc \equiv  \frac{\nu}{D_p} \sim 1 + St^2.
\end{equation}

\noindent where $\nu = \alpha H^2_g \Omega_K$ is the turbulent viscosity of the gas (assumed to be equivalent to the gas diffusivity) expressed through the non-dimensional $\alpha$ parameter measuring the level of turbulence within the disk \citep{SS73} and $St$ is the Stokes number, corresponding to the stopping time multiplied by the local Keplerian frequency. Larger values of $\alpha$ yield a more efficient redistribution of the dust grains through turbulent diffusion. Finally, the advection term in the transport equation is given by \citet{Ci10,Ci11} as:

\begin{equation}
\label{adv}
v_{adv} = \frac{D_p}{\rho_g}\frac{\partial \rho_g}{\partial z} + \frac{\partial D_p}{\partial z} + v_{d,z}.
\end{equation} 

This set of equations allows us to derive the vertical position $z_d$ of individual particles at each timestep and to subsequently estimate the dose of irradiation they received. 

\section{Irradiation of grains}

The energy received by water molecules per unit time due to cosmic rays irradiation $W_{irr}(n)$ as a function of the column density of gas $n$ above the particle is taken from \cite{Ye11}. These authors computed this term by using the cosmic ray intensity $I(E)$ (cm$^{-2}$ s$^{-1}$st$^{-1}$MeV$^{-1}$) from \cite{Co03} and considering a planar geometry. This leads to a differential flux spectrum inside the disk $F(E)$ given by:

\begin{equation}
F(E) = \pi I(E) ({\rm cm^{-2} s^{-1} MeV^{-1}}).
\end{equation}

\noindent $W_{irr}(n)$ is then derived from the molecular cloud material stopping power $S(E)$, i.e. its capability to absorb the cosmic rays energy, as estimated from the Stopping and Range of Ions in Matter (SRIM) program \citep{Zi10}. Note that the stopping power mostly depends on the mass density and not {\bf on the intrinsic composition} or structure of the material. The computation is made for the proton component of the cosmic rays and the contribution of the alpha particles is estimated approximately while the contribution of heavier cosmic rays particles is neglected. We consider that the approach of \cite{Ye11} provides the best prescription of $W_{irr}(n)$ in the literature because i) this term weakly depends on the composition, ii) the planar hypothesis is consistent with the geometry of the disk, and iii) the CRF derived from \cite{Co03} is the best proxy available for the one received by the PSN. 

At each timestep, the column density of gas above a given particle $p$ is calculated as

\begin{equation}
n = \frac{N_A}{\mu_g} \int^{+\infty}_{\mid{Z_p}\mid} \rho_g(z) dz,
\end{equation}

\noindent where $N_A$ is Avogadro's number. The energy deposited onto water molecules during the timestep $dt$ is then given by

\begin{equation}
E_{dep} = W_{irr} (n)~dt.
\end{equation}

O$_2$ is produced by radiolysis of water ice through the chemical reaction 2 H$_2$O $\rightarrow$ 2 H$_2$ + O$_2$, with an amount of energy $E_w$ needed to alter one H$_2$O molecule being $E_w$~=~235 eV \citep{Jo91}. The fraction of O$_2$ produced by the alteration of two water molecules, assuming a full efficiency, is then

\begin{equation}
\frac{[O_2]}{[H_2O]} = \frac{E_{dep}(t)}{2E_W}.
\end{equation}

{\bf \noindent Note that the value of $E_w$ = 235 eV is a lower limit, and that ``full efficiency'' assumption may be too generous. In fact \cite{Teo17} find for highly penetrating radiation (such as cosmic rays penetrating through grains) that average G-values for O$_2$ have an approximate inverse dependence on particle penetration range (see their Figure 2).}

\section{Results}

Turbulence plays an important role in the motion of gas-coupled small particles. Here, micron-sized grains initially settled in the midplane are entrained by turbulent eddies and diffuse both radially and vertically with an effective viscosity roughly equal to that of the gas \citep{Ci12}. Consequently, solid particles follow a Gaussian distribution in the vertical direction. The scale height of dust (corresponding to the standard deviation of the distribution) is a fraction of the gas scale height, this fraction being larger and possibly equal to the gas scale height $H_g$ in the cases of small grains and higher degrees of turbulence. 

\begin{figure}[h]
\centerline{\includegraphics[width=8cm]{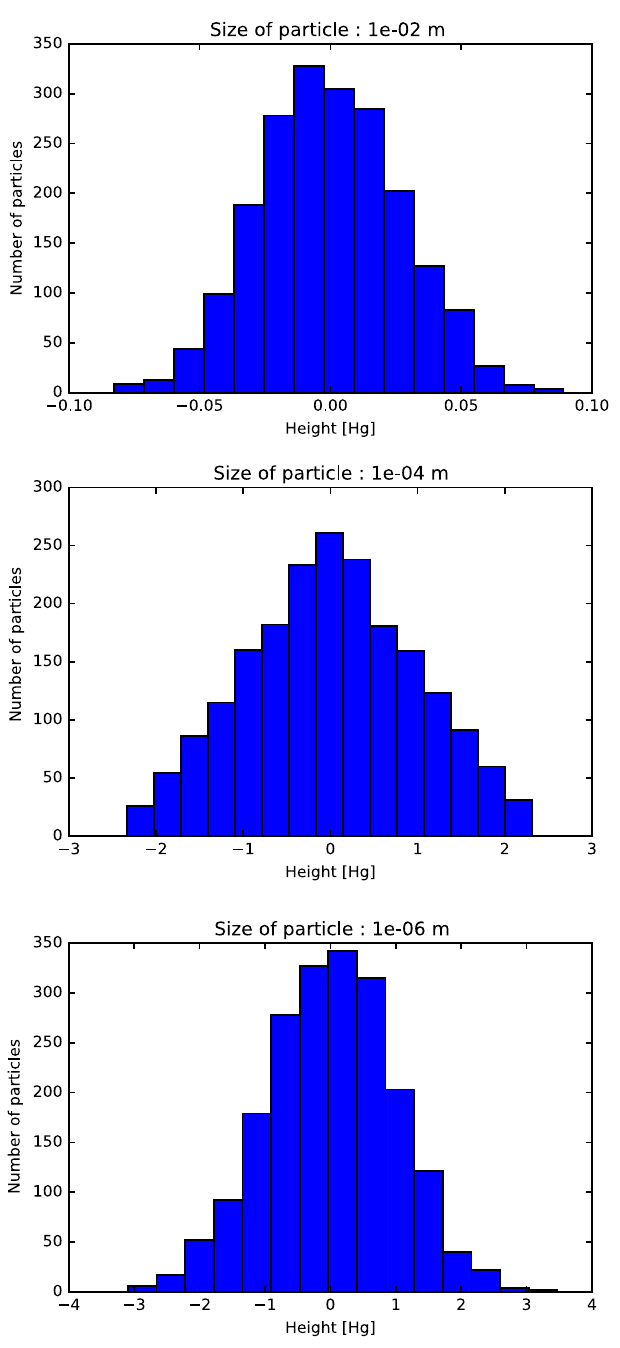}}
\caption{Vertical distribution of 10$^{-6}$, 10$^{-4}$, and 10$^{-2}$ m particles at 30 AU in the PSN. The height above the disk is expressed in function of the disk's gas scale height ($H_g$).}
\label{fig2}
\end{figure}

\begin{figure}[h]
\centerline{\includegraphics[width=\columnwidth]{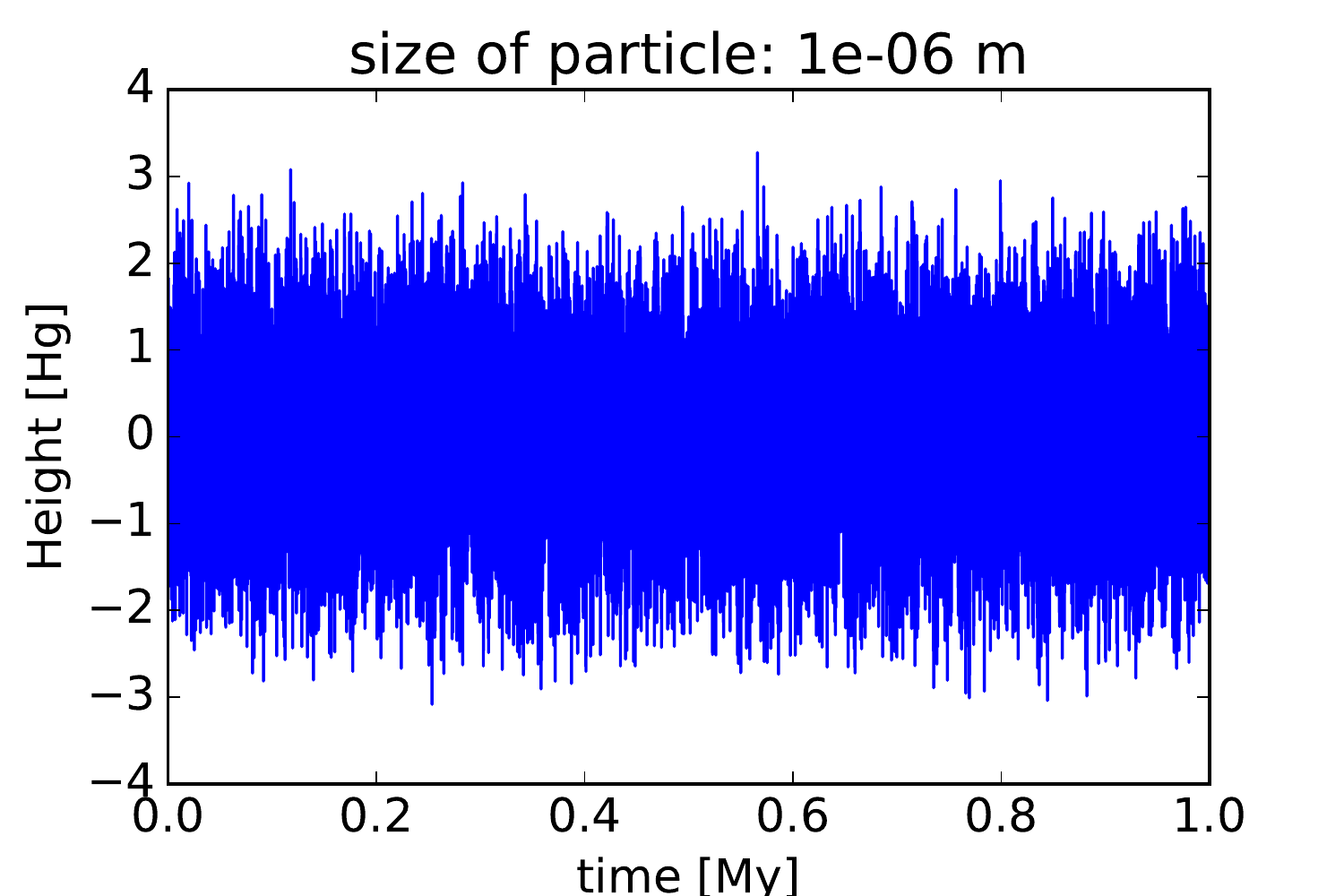}}
\caption{Vertical evolution of a 10$^{-6}$ m particle as a function of time at 30 AU in the PSN.}
\label{fig3}
\end{figure}

Figure \ref{fig2} represents the vertical distribution of 10$^{-6}$, 10$^{-4}$, and 10$^{-2}$ m particles computed at a fixed distance of 30 AU from the Sun with our disk model with vertical transport, assuming a coefficient of turbulent viscosity $\alpha$~=~10$^{-3}$. The vertical transport of 2,000 particles is simulated in each case. The figure shows that vertical spreading is more important in the cases of 10$^{-6}$ and 10$^{-4}$ m particles and can reach up to 2--3 gas scale heights. In contrast, because of their larger size, 10$^{-2}$ m particles are much less affected by turbulence and do not spread more than $\sim$0.1 scale height above the midplane. Figure \ref{fig3} displays the vertical evolution of a 10$^{-6}$ m particle integrated over 1 Myr at 30 AU in the PSN. It also shows that this particle spends a non negligible amount of time in the regions above the disk midplane, where irradiation is more significant. 10$^{-4}$ m particles display similar behaviors to a slightly lower extent while 10$^{-2}$ m particles remain mostly close to the disk midplane.

\begin{figure}[h]
\centerline{\includegraphics[width=10cm]{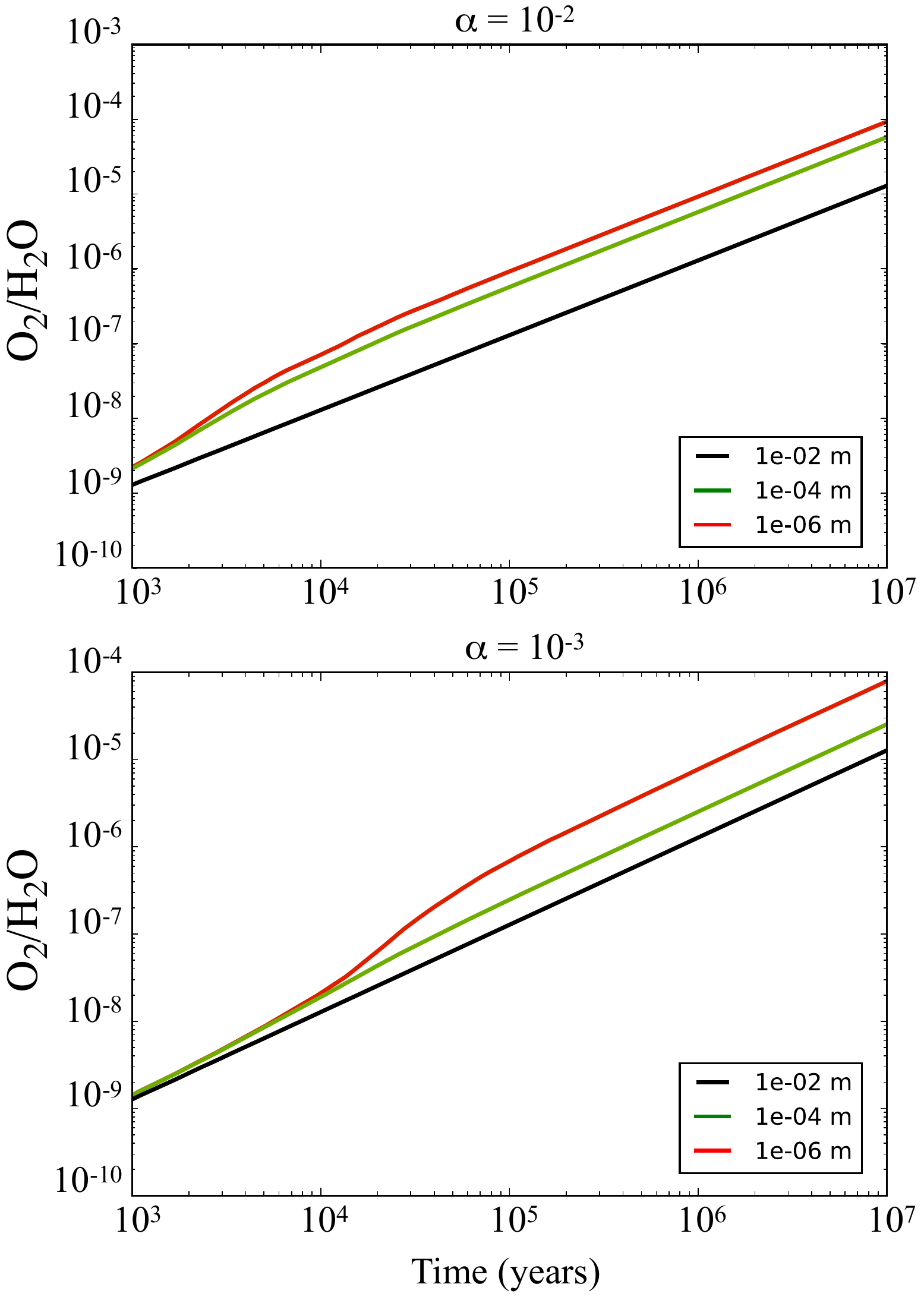}}
\caption{Abundance of O$_2$ relative to H$_2$O in 10$^{-6}$, 10$^{-4}$, and 10$^{-2}$ m particles as a function of time in the PSN and for $\alpha$ values equal to 10$^{-2}$ (top panel) and 10$^{-3}$ (bottom panel).}
\label{fig4}
\end{figure}

Figure \ref{fig4} shows the time evolution of the O$_2$/H$_2$O ratio in 10$^{-6}$, 10$^{-4}$, and 10$^{-2}$ grains irradiated along their vertical trajectories and for a turbulent viscosity $\alpha$ equal to 10$^{-3}$ and 10$^{-2}$, namely two typical values for the disk's viscosity parameter (Drouart et al. 1999). The resulting O$_2$/H$_2$O ratio in 10$^{-6}$ m particles (the most favorable case) is at best $\sim$10$^{-5}$, namely three orders of magnitude lower than the one observed in 67P/C-G by the Rosetta spacecraft, after 1 Myr of vertical transport in the PSN, irrespective of the adopted $\alpha$ value. After the same timespan, the O$_2$/H$_2$O ratio reaches no more than $\sim$10$^{-6}$ in 10$^{-2}$ m particles, while the one obtained in 10$^{-4}$ m grains is in the $\sim$2--6 $\times$ 10$^{-6}$ range, depending on the chosen value of $\alpha$. After 10 Myr of vertical transport, the O$_2$/H$_2$O ratio only increases by one order of magnitude in each considered case, leading to a O$_2$/H$_2$O ratio of $\sim$10$^{-4}$ in the most favorable situation (10$^{-6}$ m particles), which is still two orders of magnitude lower than the value observed in 67P/C-G.

\section{Discussion}

Our computations suggest that, even if a significant fraction of the icy particles have followed a back and forth cycle towards the upper layers of the disk over 10 million of years in a static PSN, the amount of O$_2$ created via radiolysis is at least $\sim$two orders of magnitude lower than the Rosetta observations. This timespan exceeds by far the formation timescale of 67P/C-G, which has been estimated to range between 2.2 and 7.7 Myr after the formation of Ca-Al-rich inclusions in the PSN \citep{Mo17b}. The O$_2$/H$_2$O ratios derived from our simulations at 1 Myr can already be considered as optimistic since the particles most likely grew and decoupled from gas after a few dozens of thousand years of the disk evolution \citep{WZ93}. In this case, once the icy grains have grown up to sizes larger than a few meters in the PSN, the bulk of the ice should remain unaltered by irradiation. Also, an increase of the CRF up to a factor of $\sim$100 due to a close supernova explosion would not substantially change the O$_2$/H$_2$O ratio in icy grains because the timespan of such an event (a few kyr) is too short \citep{Mo16}.

Several alternative mechanisms have been recently investigated in the literature to account for the O$_2$ detection in 67P/C-G. Among them, \cite{Ta16} have proposed that O$_2$ could be produced in dark clouds via a combination of gas-phase and solid-state chemical reactions leading to its formation and destruction, in agreement with the ROSINA observations and the conclusions of \cite{Mo16}. Another mechanism, proposed by \cite{Du17}, consists in the production of O$_2$ through dismutation of H$_2$O$_2$ during water ice desorption from the nucleus. However, this mechanism requires the incorporation of large amounts of primordial H$_2$O$_2$ into the nucleus and its complete conversion into O$_2$ to be consistent with the low levels of H$_2$O$_2$ observed in the coma. Another scenario investigated by \cite{Ya17} is the possible present-day production of O$_2$ via an Eley-Rideal reaction mechanism in the coma. This reaction between energetic water ions and adsorbed O-atoms, produces highly excited oxywater (H$_2$O$_2$), which undergoes delayed fragmentation to form HO$_2$ as the precursor for O$_2$. However, at close distance to the Sun, the solar wind strengthens and increases the ionization and water ion flux to the surface of the nucleus, and consequently produces more O$_2$ if the Eley-Rideal reaction mechanism is effective. According to this mechanism, the O$_2$/H$_2$O ratio should increase at perihelion, a trend that has not been seen by the ROSINA instrument.

Our results thus favor the mechanism of radiolysis of icy grains in low-density environments such as the presolar cloud, shown by \cite{Mo16} to be capable of producing the O$_2$ abundance observed in 67P/C-G. {\bf They suggested that O$_2$ may be trapped in the grains in radiation defects/cavities, and subsequently delivered to the PSN either in solid or gas phase, depending on the disk's thermodynamic structure.} The constant O$_2$/H$_2$O ratio observed in 67P/C-G requires that both O$_2$ and H$_2$O were released from the same solid phase and supports the hypothesis suggesting that comets formed from water ice coming from ISM without suffering from vaporization when entering the PSN\footnote{This scenario does not preclude a possible amorphous-to-crystalline ice phase transition due to a moderate PSN temperature along the migration path the grains \citep{Mo16}.}. Relaxing this {\bf constraint} would leave alternative delivery scenarios of O$_2$ to comets, among which {\bf the desorption of this molecule during the amorphous-to-crystalline ice phase transition encompassed by presolar grains that entered into the disk. The O$_2$ released via this manner could have been subsequently trapped in clathrates during the cooling of the PSN} before being incorporated in the building blocks of comets. Finally, we note that, in addition to O$_2$ formation in low-density environments at epochs prior to the genesis of the protosolar nebula, an endogenic radiolytic source due to radionuclides present in the dusty component of the comet nucleus may have contributed at the percent level to the total O$_2$ budget in comets \citep{Bo17}.

\acknowledgements
We thank an anonymous Referee for his helpful comments. O.M. acknowledges support from CNES. O.M. and T.R. acknowledge support from the A*MIDEX project (n\textsuperscript{o} ANR-11-IDEX-0001-02) funded by the ``Investissements d'Avenir'' French Government program, managed by the French National Research Agency (ANR). J.I.L. was supported by the JWST project through a grant from NASA/GSFC.



\newpage 

\newpage 

\newpage 

\end{document}